\documentclass{article}
\usepackage{templates/ijcai24}

\usepackage{times}
\usepackage{soul}
\usepackage{url}
\usepackage[hidelinks,breaklinks]{hyperref}
\usepackage[utf8]{inputenc}
\usepackage[small]{caption}
\usepackage{graphicx}
\usepackage{amsmath}
\usepackage{amsthm}
\usepackage{booktabs}
\usepackage{algorithm}
\usepackage{algorithmic}
\usepackage[switch]{lineno}

\usepackage{xcolor}
\usepackage{xspace}
\usepackage{enumitem}
\usepackage{wrapfig}

\definecolor{red3}{HTML}{f03c1d}
\definecolor{blue3}{HTML}{2e4da3}
\definecolor{yellow3}{HTML}{fca00a}

\newcommand{\de}[0]{Diffusion Explainer\xspace}
\newcommand{\sd}[0]{Stable Diffusion\xspace}

\title{Interactive Visual Learning for Stable Diffusion}

\author{
Seongmin Lee$^1$
\and
Benjamin Hoover$^{1,2}$\and
Hendrik Strobelt$^{2}$\and
Zijie J. Wang$^{1}$\and\\
ShengYun Peng$^{1}$\and
Austin Wright$^{1}$\and
Kevin Li$^{1}$\and
Haekyu Park$^{1}$\and\\
Haoyang Yang$^{1}$\And
Polo Chau$^1$\\
\affiliations
$^1$Georgia Tech\\
$^2$IBM Research\\
\emails
\{seongmin, bhoov\}@gatech.edu,
hendrik.strobelt@ibm.com,
\{jayw,speng65,apwright,kevin.li,haekyu,alexanderyang,polo\}@gatech.edu
}

\begin{document}

\maketitle

\begin{abstract}
Diffusion-based generative models' impressive ability to create convincing images has garnered global attention.
However, their complex internal structures and operations often pose challenges for non-experts to grasp.
We introduce \de{}, the first interactive visualization tool designed to elucidate  how \sd transforms text prompts into images.
It tightly integrates a visual overview of Stable Diffusion’s complex components with detailed explanations of their underlying operations.
This integration enables users to fluidly transition between multiple levels of abstraction through animations and interactive elements.
Offering real-time hands-on experience,
\de allows users to adjust \sd's hyperparameters and prompts without the need for installation or specialized hardware.
Accessible via users' web browsers, 
\de is making significant strides in democratizing AI education, 
fostering broader public access.
More than 7,200 users spanning 113 countries have used our
open-sourced tool at \url{https://poloclub.github.io/diffusion-explainer/}.
A video demo is available at \url{https://youtu.be/MbkIADZjPnA}. %
\end{abstract}

\section{Introduction}
\label{sec:intro}
\begin{figure*}
    \centering
    \includegraphics[width=\textwidth]{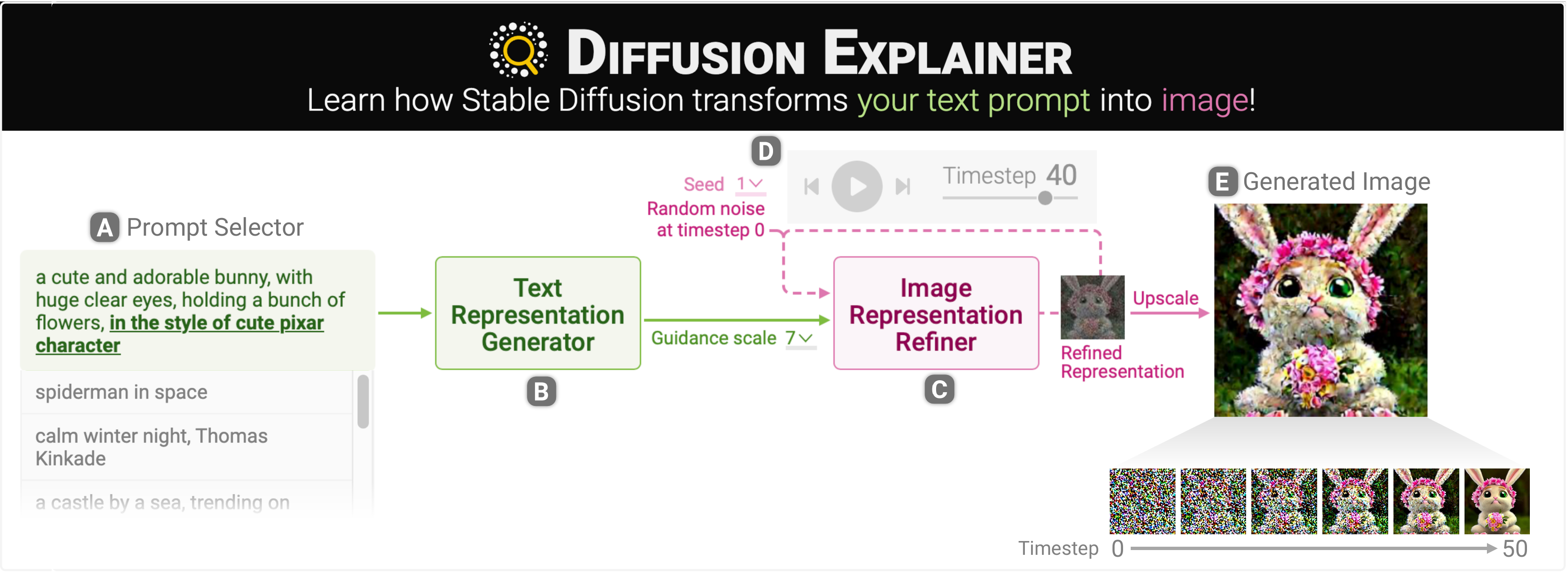}
    \caption{With \de, users can  examine how
    \textbf{(A)} \textit{a text prompt}, e.g., \textit{``a cute and adorable bunny... pixar character''}, is encoded by 
    \textbf{(B)} the \textit{Text Representation Generator} into vectors to guide 
    \textbf{(C)} the \textit{Image Representation Refiner} to iteratively refine the vector representation of the image being generated.
    \textbf{(D)} The \textit{Timestep Controller} enables users to review the incremental improvements in image quality and adherence to the prompt over timesteps. 
    \textbf{(E)} The final image representation is upscaled to a high-resolution image.
    Diffusion Explainer tightly integrates a visual overview of \sd's complex components with detailed explanations, 
    enabling users to fluidly transition between 
    abstraction levels through animations and interactive elements (see \autoref{fig:text_expand} and \autoref{fig:image_expand}).}
    \label{fig:architecture}
\end{figure*}

Diffusion-based generative models~\cite{rombach2022high}
like Stable Diffusion~\cite{stability2022stable} 
have captured global attention for their impressive image creation abilities, from AI developers, designers, to policymakers.
However, 
the popularity and progress of generative AI models have sparked ethical~\cite{brusseau2022acceleration} and social concerns,
such as accusations of artistic style theft by AI image generators~\cite{sung2022lensa,choudhary2022stable}.
Policymakers are also discussing ways to combat malicious data generation and revise copyright policies~\cite{engler2023early,ryanmosley2023an,2023copyright}.
There is an urgent need for individuals from many different fields to understand how generative AI models function and communicate effectively with AI researchers and developers~\cite{dixit2023meet,hendrix2023generative}. 

\smallskip
\noindent
\textbf{Key challenges in designing learning tools for \sd.}
\sd iteratively refines \textit{noise} into a high-resolution image's vector representation, guided by a text prompt. 
Internally, the prompt is tokenized and encoded into vector representations by the \textit{CLIP}'s \textit{Text Encoder}~\cite{radford2021learning}.
With text representations' guidance,
\sd improves the image quality and adherence to the prompt by incrementally denoising the image's vector representation using the \textit{UNet}~\cite{ronneberger2015u} and the \textit{Scheduler} algorithm~\cite{nichol2021improved}.
The final image representation is upscaled to a high-resolution image.
The crux of learning about \sd tems from the
complex interplay between the multiple neural network subcomponents, their intricate operations, and the iterative nature of image representation refinements.
Such complex interactions are challenging even for experts to comprehend~\cite{platen2022testing}. 
While some articles~\cite{alammar2022the} and video lessons~\cite{howard2023from} explain \sd,
they 
often focus on model training and mathematical details. 

\medskip
\noindent
\textbf{Contributions.} In this demonstration, we contribute: 

\begin{itemize}[topsep=0pt, itemsep=0mm, parsep=3pt, leftmargin=10pt]
    \item \textbf{\de,
    the first interactive visualization tool designed for non-experts}
    to learn how \sd transforms a text prompt into a high-resolution image (\autoref{fig:architecture}), overcoming design challenges in developing learning tools for \sd. 
    \de integrates an overview of \sd's complex structure with explanations of their underlying operations 
    enabling users to fluidly transition between multiple abstraction levels through animations and interactive elements. %
    \item \textbf{Real-time interactive visualization}
    to discover how \sd's hyperparameters and text prompt affect image generation,
    empowering users to experiment with their settings and gain insight into each hyperparameter's impact without the need for complex mathematical derivations.
    \item \textbf{Open-sourced, web-based implementation}
    that broadens the public's education access to modern generative AI
    without requiring any installation, advanced computational resources, or coding skills.
    \de is open-sourced\footnote{\url{https://github.com/poloclub/diffusion-explainer}}
    and available at \url{https://poloclub.github.io/diffusion-explainer/}.
    A video demo is available at \url{https://youtu.be/MbkIADZjPnA}. %
With over 7,200 users across 113 countries, \de is making significant strides in democratizing AI education.
\end{itemize}

\section{System Design and Implementation}
\label{sec:design}

\de is an interactive visualization tool that explains how \sd generates a high-resolution image from a text prompt.
It incorporates an animation of random noise gradually refined and a \textit{Timestep Controller} (\autoref{fig:architecture}D) that enables users to visit each refinement timestep.
From the \textit{Prompt Selector} (\autoref{fig:architecture}A),
users select one out of the 13 prompts that follow a template and contain popular keywords
identified from literature~\cite{smith2022traveler}.
\de provides an overview of \sd's architecture, which can be expanded into details via user interactions (\autoref{fig:text_expand}, \autoref{fig:image_expand}).
While users can interactively change \sd's two key hyperparameters, guidance scale and random seed, 
we fix the number of timesteps as 50, a commonly chosen value,
and consistently use the Linear Multistep Scheduler~\cite{karras2022elucidating}, a fundamental and widely adopted scheduling method.
\de is implemented using a standard web technology stack (HTML, CSS, JavaScript) and the D3.js~\cite{bostock2011d3} visualization library.

\subsection{Text Representation Generator}
\label{sec:text}
\begin{figure}[t]
    \centering
    \includegraphics[width=\columnwidth]{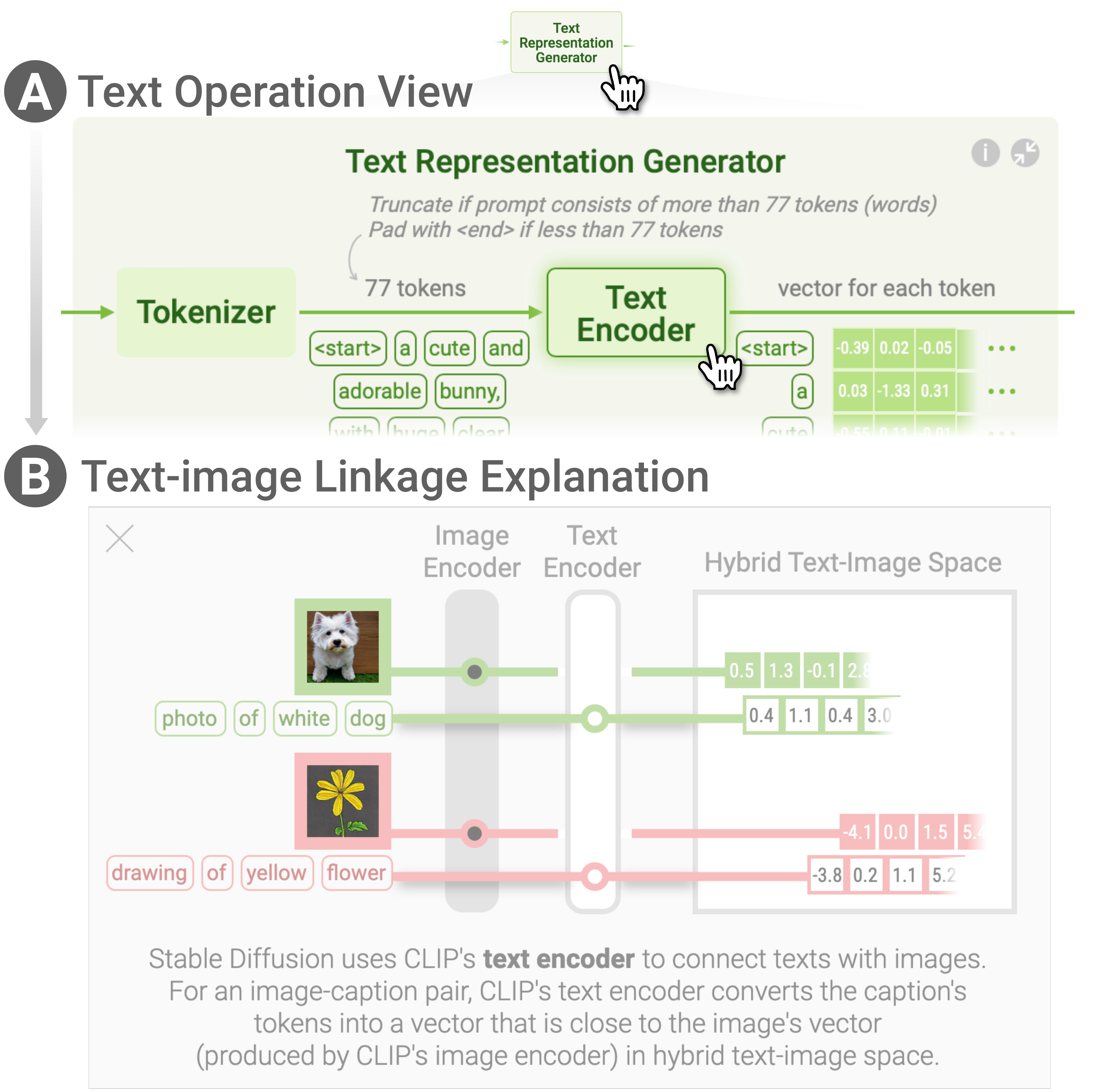}
    \caption{
    To understand how \sd converts a text prompt into vector representations, users click on the \textit{Text Representation Generator}, which smoothly expands to
    \textbf{(A)} the \textit{Text Operation View} that explains how the prompt is split into tokens and encoded into vector representations.
    \textbf{(B)} The \textit{Text-image Linkage Explanation}
    demonstrates how \sd bridges text and image, 
    enabling text representations to guide the image generation process.
    }
    \label{fig:text_expand}
\end{figure}

The \textit{Text Representation Generator} converts text prompts into vector representations.
Clicking on the \textit{Text Representation Generator} 
expands to the \textit{Text Operation View} (\autoref{fig:text_expand}A) that explains how the Tokenizer splits the prompt into tokens
and how the Text Encoder encodes the tokens into vector representations.
Clicking on the Text Encoder displays the \textit{Text-image Linkage Explanation} (\autoref{fig:text_expand}B), 
which visually explains how \sd connects text and image by utilizing the CLIP~\cite{radford2021learning} text encoder to generate text representations with image-related information.

\subsection{Image Representation Refiner}
\label{sec:image}
\begin{figure}[t]
    \centering
    \includegraphics[width=\columnwidth]{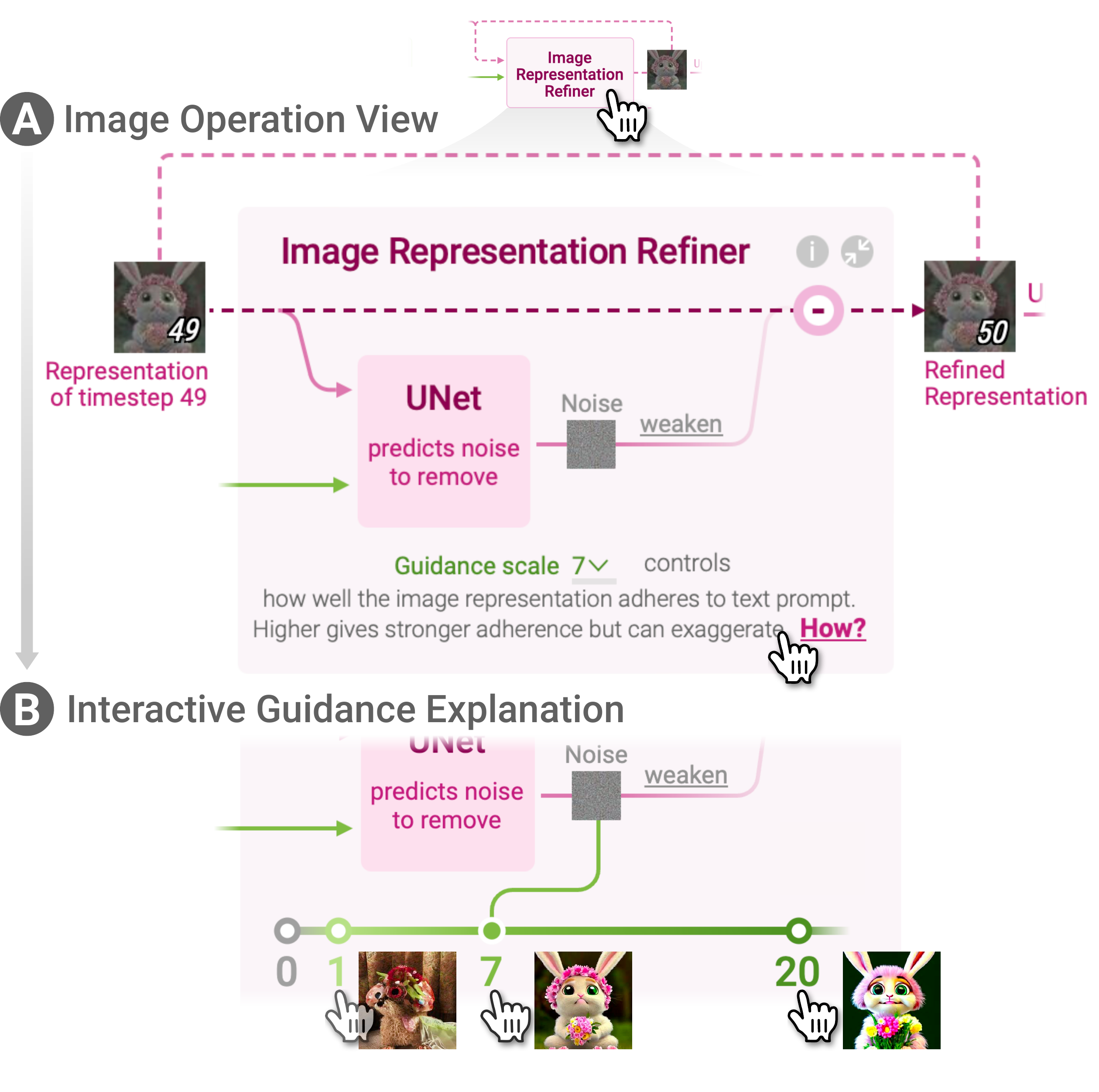}
    \caption{Users learn how \sd gradually refines noise into a high-resolution image's vector representation aligned with the text prompt
    by selecting the \textit{Image Representation Refiner} from the high-level overview.
    This smoothly expands to 
    \textbf{(A)} the \textit{Image Operation View}
    that demonstrates how
    the noise is iteratively %
    predicted and removed from the image representation. %
    \textbf{(B)} The \textit{Interactive Guidance Explanation} enables users to interactively experiment with various guidance scale values (0, 1, 7, 20) 
    to better understand how higher values lead to stronger adherence.} %
    \label{fig:image_expand}
\end{figure}

The \textit{Image Representation Refiner} (\autoref{fig:image_expand}) refines random noise into the vector representation of a high-resolution image that adheres to the text prompt.
\de visualizes the image representation of each refinement step in two ways:
(1) decoding it as a small image using linear operations~\cite{turner2022decoding} and 
(2) upscaling it to the \sd's output resolution (\autoref{fig:architecture}E).
Users expands the Image Representation Refiner to access the \textit{Image Operation View} (\autoref{fig:image_expand}A),
which explains how the UNet neural network~\cite{ronneberger2015u} predicts the noise to be removed from the image representation. %

The guidance scale hyperparameter, which controls the image's adherence strength to the text prompt, is described at the bottom, and further explained in the \textit{Interactive Guidance Explanation}  (\autoref{fig:image_expand}B). Using a slider, users can experiment with different guidance scale values to better understand how higher values lead to stronger adherence of the generated image to the text prompt.

\section{Demonstrating \de}
\label{sec:scenario}
We provide a demonstration scenario
both to illustrate how people with limited experience with \sd 
may benefit from using \de and to describe what we will show the audience.
\subsubsection{Demonstration Scenario}
Troy, a government policymaker
overseeing AI image creation in the entertainment and media industries,
has recently received concerns from artists. 
They express worry that their artwork has been exploited by AI models to create commercial products without their consent~\cite{amelion2023twitter}.
Troy is eager to help these artists 
in getting compensated for their contributions.
He has found a tool
that could potentially address their concerns, which would attribute AI-generated images to human artists~\cite{huber2023stable,anton2023twitter}.
However, before proposing any policies,
Troy needs to understand how and if such attribution may work.

Troy launches \de
which illustrates how \sd transforms a text prompt into a high-resolution image through an iterative process (\autoref{fig:architecture}).
He identifies two controllable hyperparameters: \textit{random seed} and \textit{guidance scale}. 
Adjusting the random seed from 1 to 2 and 3,
Troy observes 
substantial changes in the generated image.
Intrigued by these variations, 
he examines timestep~1 
using the \textit{Timestep Controller} (\autoref{fig:architecture}D) 
and discovers that different random seeds yield different initial noises, thus generating diverse images.
Continuing his exploration, 
Troy experiments with different guidance scale values.
He notes that a guidance scale value of 7
produces a realistic image closely aligned with the text prompt,
while values of 1 or 20 result in images that are hard to interpret or exaggerated.
To delve into the %
details of how the text prompt is processed,
Troy clicks on the \textit{Text Representation Generator} to expand it into the \textit{Text Operation View}  (\autoref{fig:text_expand}A). 
Here, he discovers that the prompt is tokenized and 
converted into vector representations.
Seeking clarity on how text is connected to the image,
he then displays the \textit{Text-image Linkage Explanation}  (\autoref{fig:text_expand}B)
and learns that \sd's 
text representations contain image-related information.
Troy proceeds to 
understand %
the refinement %
of image representation by examining the \textit{Image Operation View}  (\autoref{fig:image_expand}A).
He discovers that each refinement step involves 
UNet's noise prediction and removal,
with the guidance scale hyperparameter controlling the adherence of the generated image to the prompt.
Intrigued,
Troy accesses the \textit{Interactive Guidance Explanation}  (\autoref{fig:image_expand}B)
and learns that 
the model predicts two types of noise, each of which is generic and prompt-specific. 
The final noise is a weighted sum of these noises, with the weight being controlled by the guidance scale.
With an improved understanding of the image generation process of \sd, Troy recognizes that 
image analysis alone, without considering text prompts, 
will not suffice to discern how an artist's creations contributed to AI-generated images.
He asserts that further research is imperative to accurately attribute AI-generated images.

\section{Conclusion}
\label{sec:conclusion}
\de, the first interactive visualization for non-experts, explains how \sd generates high-resolution images from text prompts. 
It tightly integrates a visual overview of \sd's 
components with detailed explanations of their underlying operations
and
provides real-time hands-on experience to change \sd's hyperparameters and prompts
on the browser without any installation or hardware requirements.
More than 7,200 users spanning 113 countries have used \de.
\bibliographystyle{templates/named}
\bibliography{references}

\begin{thebibliography}{}

\bibitem[\protect\citeauthoryear{Alammar}{2022}]{alammar2022the}
Jay Alammar.
\newblock {The illustrated Stable Diffusion}.
\newblock \url{https://jalammar.github.io/illustrated-stable-diffusion/}, 2022.
\newblock Accessed on: 2023-04-30.

\bibitem[\protect\citeauthoryear{{AMELION}}{2023}]{amelion2023twitter}
{AMELION}.
\newblock \url{https://twitter.com/amelion_/status/1651193228677218304}, 2023.
\newblock Accessed on: 2023-04-26.

\bibitem[\protect\citeauthoryear{{anton}}{2023}]{anton2023twitter}
{anton}.
\newblock {Announcing Stable Attribution - A tool which lets anyone find the
  human creators behind AI generated images.}
\newblock \url{https://twitter.com/atroyn/status/1622355473193381888}, 2023.
\newblock Accessed on: 2023-04-30.

\bibitem[\protect\citeauthoryear{Bostock \bgroup \em et al.\egroup
  }{2011}]{bostock2011d3}
Michael Bostock, Vadim Ogievetsky, and Jeffrey Heer.
\newblock {D$^3$ Data-driven Documents}.
\newblock {\em IEEE transactions on visualization and computer graphics},
  17(12):2301--2309, 2011.

\bibitem[\protect\citeauthoryear{Brusseau}{2022}]{brusseau2022acceleration}
James Brusseau.
\newblock {Acceleration AI Ethics, the Debate between Innovation and Safety,
  and Stability AI's Diffusion versus OpenAI's Dall-E}.
\newblock {\em arXiv preprint arXiv:2212.01834}, 2022.

\bibitem[\protect\citeauthoryear{Choudhary}{2022}]{choudhary2022stable}
Lokesh Choudhary.
\newblock {Stable Diffusion is Now Accused of ‘Stealing’ Artwork}.
\newblock
  \url{https://analyticsindiamag.com/stable-diffusion-is-now-accused-of-stealing-artwork/},
  2022.
\newblock Accessed on: 2023-04-30.

\bibitem[\protect\citeauthoryear{Dixit}{2023}]{dixit2023meet}
Pranav Dixit.
\newblock {Meet The Three Artists Behind A Landmark Lawsuit Against AI Art
  Generators}.
\newblock
  \url{https://www.buzzfeednews.com/article/pranavdixit/ai-art-generators-lawsuit-stable-diffusion-midjourney},
  2023.
\newblock Accessed on: 2023-04-30.

\bibitem[\protect\citeauthoryear{Engler}{2023}]{engler2023early}
Alex Engler.
\newblock {Early thoughts on regulating generative AI like ChatGPT}.
\newblock {\em Brookings Institution}, 2023.
\newblock Accessed on: 2023-04-30.

\bibitem[\protect\citeauthoryear{Hendrix}{2023}]{hendrix2023generative}
Justin Hendrix.
\newblock {Generative AI, Section 230 and Liability: Assessing the Questions}.
\newblock {\em Tech Policy Press}, 2023.
\newblock Accessed on: 2023-04-30.

\bibitem[\protect\citeauthoryear{Howard}{2023}]{howard2023from}
Jeremy Howard.
\newblock {From Deep Learning Foundations to Stable Diffusion}.
\newblock \url{https://www.fast.ai/posts/part2-2023.html}, 2023.
\newblock Accessed on: 2023-04-30.

\bibitem[\protect\citeauthoryear{Huber and Troynikov}{2023}]{huber2023stable}
Jeff Huber and Anton Troynikov.
\newblock {Stable Attribution}.
\newblock \url{https://www.stableattribution.com}, 2023.
\newblock Accessed on: 2023-04-30.

\bibitem[\protect\citeauthoryear{Karras \bgroup \em et al.\egroup
  }{2022}]{karras2022elucidating}
Tero Karras, Miika Aittala, Timo Aila, and Samuli Laine.
\newblock Elucidating the design space of diffusion-based generative models.
\newblock {\em Advances in Neural Information Processing Systems},
  35:26565--26577, 2022.

\bibitem[\protect\citeauthoryear{Nichol and
  Dhariwal}{2021}]{nichol2021improved}
Alexander~Quinn Nichol and Prafulla Dhariwal.
\newblock {Improved Denoising Diffusion Probabilistic Models}.
\newblock In {\em International Conference on Machine Learning}, pages
  8162--8171. PMLR, 2021.

\bibitem[\protect\citeauthoryear{Radford \bgroup \em et al.\egroup
  }{2021}]{radford2021learning}
Alec Radford, Jong~Wook Kim, Chris Hallacy, Aditya Ramesh, Gabriel Goh,
  Sandhini Agarwal, Girish Sastry, Amanda Askell, Pamela Mishkin, Jack Clark,
  et~al.
\newblock {Learning Transferable Visual Models from Natural Language
  Supervision}.
\newblock In {\em International conference on machine learning}, pages
  8748--8763. PMLR, 2021.

\bibitem[\protect\citeauthoryear{Rombach \bgroup \em et al.\egroup
  }{2022}]{rombach2022high}
Robin Rombach, Andreas Blattmann, Dominik Lorenz, Patrick Esser, and Bj{\"o}rn
  Ommer.
\newblock {High-resolution Image Synthesis with Latent Diffusion Models}.
\newblock In {\em Proceedings of the IEEE/CVF Conference on Computer Vision and
  Pattern Recognition}, pages 10684--10695, 2022.

\bibitem[\protect\citeauthoryear{Ronneberger \bgroup \em et al.\egroup
  }{2015}]{ronneberger2015u}
Olaf Ronneberger, Philipp Fischer, and Thomas Brox.
\newblock {U-net: Convolutional Networks for Biomedical Image Segmentation}.
\newblock In {\em Medical Image Computing and Computer-Assisted
  Intervention--MICCAI 2015: 18th International Conference, Munich, Germany,
  October 5-9, 2015, Proceedings, Part III 18}, pages 234--241. Springer, 2015.

\bibitem[\protect\citeauthoryear{Ryan-Mosley}{2023}]{ryanmosley2023an}
Tate Ryan-Mosley.
\newblock {An early guide to policymaking on generative AI}.
\newblock {\em MIT Technology Review}, 2023.
\newblock Accessed on: 2023-04-30.

\bibitem[\protect\citeauthoryear{Smith}{2022}]{smith2022traveler}
Ethan Smith.
\newblock {A Traveler's Guide to the Latent Space}.
\newblock
  \url{https://sweet-hall-e72.notion.site/A-Traveler-s-Guide-to-the-Latent-Space-85efba7e5e6a40e5bd3cae980f30235f#976ba690a0904431aac693d59830a92c},
  2022.
\newblock Accessed on: 2023-04-29.

\bibitem[\protect\citeauthoryear{{Stability AI}}{2022}]{stability2022stable}
{Stability AI}.
\newblock {Stable Diffusion Public Release}.
\newblock \url{https://stability.ai/blog/stable-diffusion-public-release},
  2022.
\newblock Accessed on: 2022-08-22.

\bibitem[\protect\citeauthoryear{Sung}{2022}]{sung2022lensa}
Morgan Sung.
\newblock {Lensa, the AI portrait app, has soared in popularity. But many
  artists question the ethics of AI art.}
\newblock {\em NBC News}, 2022.
\newblock Accessed on: 2023-04-30.

\bibitem[\protect\citeauthoryear{Turner}{2022}]{turner2022decoding}
Kevin Turner.
\newblock {Decoding latents to RGB without upscaling}.
\newblock
  \url{https://discuss.huggingface.co/t/decoding-latents-to-rgb-without-upscaling/23204/2},
  2022.
\newblock Accessed on: 2023-04-30.

\bibitem[\protect\citeauthoryear{{U.S. Copyright Office}}{2023}]{2023copyright}
{U.S. Copyright Office}.
\newblock {Copyright Office Launches New Artificial Intelligence Initiative}.
\newblock \url{https://www.copyright.gov/newsnet/2023/1004.html}, 2023.
\newblock Accessed on: 2023-04-30.

\bibitem[\protect\citeauthoryear{von Platen}{2022}]{platen2022testing}
Patrick von Platen.
\newblock {Testing Stable Diffusion is hard}.
\newblock \url{https://github.com/huggingface/diffusers/issues/937}, 2022.
\newblock Accessed on: 2023-04-30.

\end{thebibliography}

\end{document}